\pdfoutput=1
\documentclass[10pt,preprintnumbers,showpacs,amsmath,amssymb,floatfix,twocolumn,prl]{revtex4-1} 
\usepackage{graphicx}
\usepackage{dcolumn}
\usepackage{bm}
\usepackage{longtable}
\usepackage{graphics}
\usepackage{amssymb}
\usepackage{amsmath}
\usepackage{xspace}
\usepackage{epsfig}
\usepackage{rotating}
\usepackage{epstopdf}
\newcommand {\ibid}{{\it ibid}. }
\newcommand {\etal}{{\it et al}. }

\newcommand {\etalc}{{\it et al}., }
\newcommand {\dmit}{Pd(dmit)$_2$ }
\newcommand {\dmitn}{Pd(dmit)$_2$}
\begin{document}
\title{Geometrical frustration in the spin liquid $\beta'$-Me$_3$EtSb[Pd(dmit)$_2$]$_2$ and the valence bond solid Me$_3$EtP[Pd(dmit)$_2$]$_2$}
\author{E. P. Scriven}
\affiliation{Centre for Organic Photonics \& Electronics, School of Mathematics \& Physics,
University of Queensland, Brisbane, Queensland 4072,
Australia}
\author{B. J. Powell}
\email{bjpowell@gmail.com}
\affiliation{Centre for Organic Photonics \& Electronics, School of Mathematics \& Physics,
University of Queensland, Brisbane, Queensland 4072,
Australia}
\pacs{}

\begin{abstract}
We show that the electronic structures of the title compounds predicted by density functional theory (DFT) are well described by tight binding models. We determine the frustration ratio, $J'/J$, of the Heisenberg model on the anisotropic triangular lattice, which describes the spin degrees of freedom in the Mott insulating phase for a range of \dmit salts. All of the antiferromagnetic materials studied have $J'/J\lesssim0.5$ or $J'/J\gtrsim0.9$ and all salts with $0.5\lesssim J'/J\lesssim0.9$ are known, experimentally, to be charge ordered, valence bond solids or spin liquids. 
\end{abstract}

\maketitle

The interplay of geometrical frustration and electronic correlations produces a wide range of exotic phenomena \cite{KanodaKato,RPP} in the organic charge transfer salts Me$_{4-n}$Et$_nPn$[Pd(dmit)$_2$]$_2$ (henceforth $Pn$-$n$) \cite{defs}. At ambient pressure and low temperature these materials are Mott insulators, many of which are driven superconducting by the application of hydrostatic pressure or uniaxial stress. Most salts display antiferromagnetic (AFM) order, but recent experiments \cite{KanodaKato,RPP} suggest that Me$_3$EtP[Pd(dmit)$_2$]$_2$ (P-1) is a valence bond solid (VBS) and Me$_3$EtSb[Pd(dmit)$_2$]$_2$ (Sb-1) is a type II spin liquid (SL) \cite{RPP}; with a singlet gap, but no triplet gap \cite{Normand}.

In this Letter we report DFT calculations of the electronic structures of Sb-1 and P-1. We parameterize these results in terms of tight binding models and report the parameters found for a number of Pd(dmit)$_2$ salts with AFM or charge ordered (CO) ground states. The simplest model that has been proposed for the insulating phases of the Pd(dmit)$_2$ salts is the Heisenberg model on the anisotropic triangular lattice \cite{KanodaKato,RPP,ShimizuJPCM}. In this model each site represents a Pd(dmit)$_2$ dimer, $J$ is the exchange coupling around the sides of square and $J'$ is the exchange interaction along one diagonal. We find that those materials that display long range AFM order lie in the parameter regimes $J'/J\lesssim0.5$ or $J'/J\gtrsim1$ where many-body theories predict long range magnetic order. Further, all of the materials with CO, SL or VBS ground states lie in the parameter regime $0.5\lesssim J'/J\lesssim0.9$ where the low energy physics remains controversial because there are a number of competing  states. We argue that this means that other terms in the Hamiltonian may be crucial for determining the ground state. 

The Heisenberg model on the anisotropic triangular lattice has been studied by a range of theoretical methods including linear spin-wave theory \cite{spinfluc},
series expansions \cite{series}, the coupled cluster method \cite{Bishop},
large-N expansions \cite{ChungJPCM01},
variational Monte Carlo \cite{Monte}, resonating valence bond theory \cite{RVB,RVB1,RVB2,RVB3,RVB4}, 
pseudo-fermion functional renormalization group \cite{Reuther10}, slave rotor theory \cite{Rao}, renormalisation group \cite{Starykh}, 
and the density matrix renormalisation group \cite{WengPRB06}.
Collectively these studies suggest that N\'eel $(\pi,\pi)$ order is realised for $J'/J\lesssim0.5$ and incommensurate $(q,q)$ long range AFM order is realised for $J'/J\sim1$ (in the special case $J'/J=1$ the 120$^\textrm{o}$ state with $q=2\pi/3$ is realised). However, the ground state for $0.6\lesssim J'/J\lesssim0.9$ remains controversial.

\begin{figure}
\begin{center}
\includegraphics[width=0.49\columnwidth]{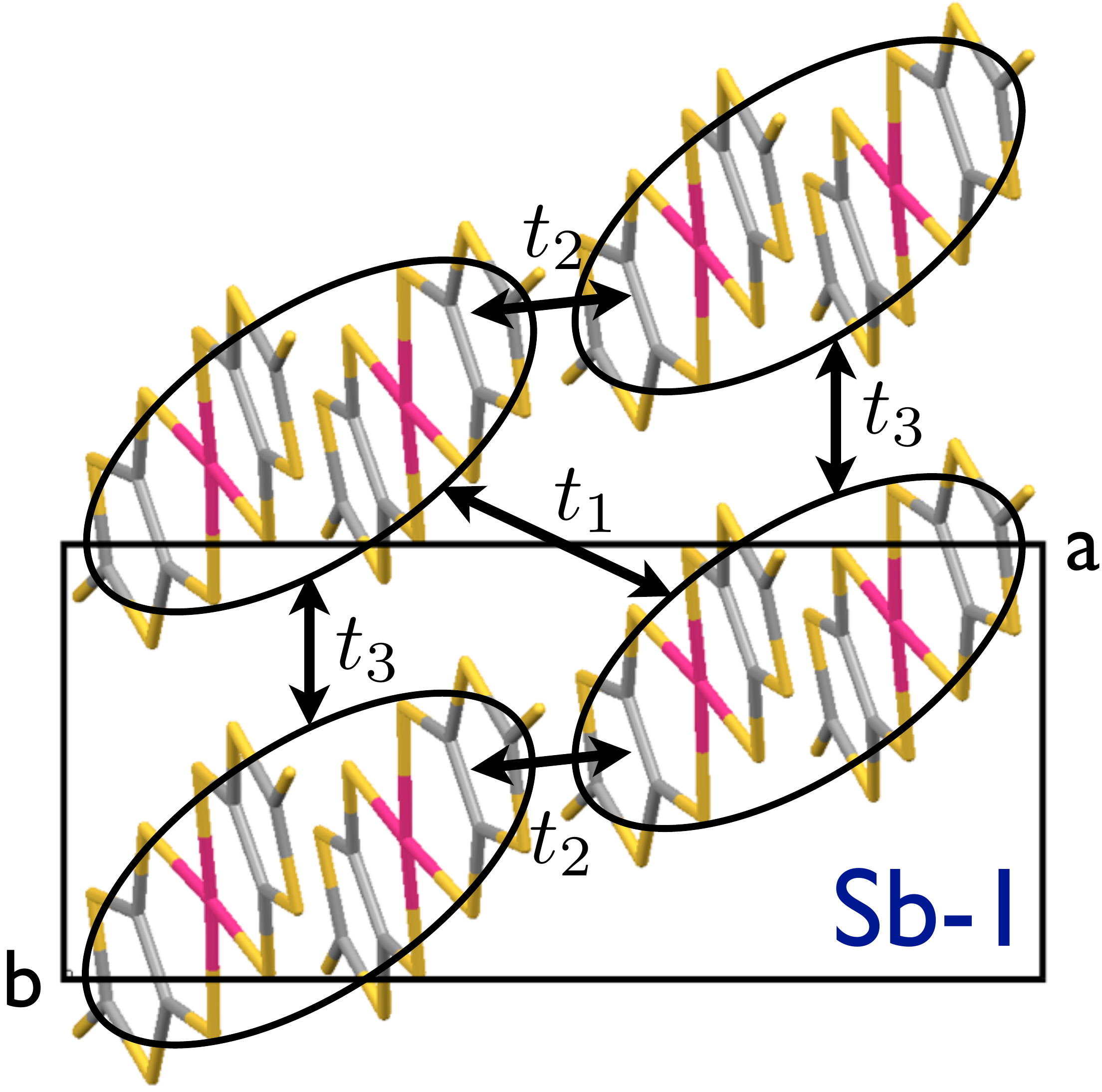}
\includegraphics[width=0.49\columnwidth]{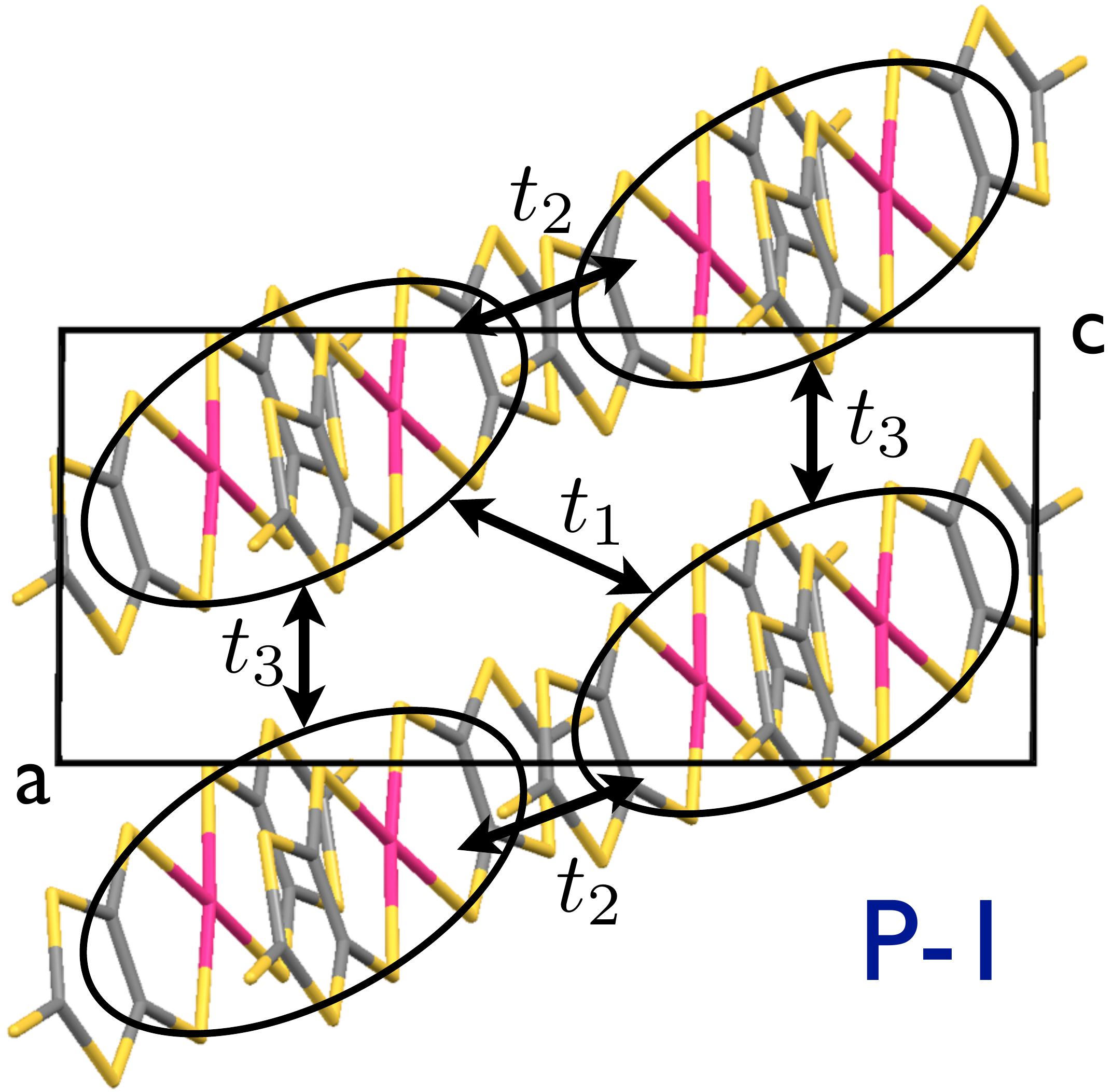}
\end{center}
\caption{(Color online.) In-plane crystal structures and inter-dimer hopping integrals for Sb-1 (left) and P-1 (right). Rings indicate the dimers. The \emph{conventional} unit cell is marked.}\label{fig:lattice}
\end{figure}

Many \dmit salts undergo a Mott transition under hydrostatic pressure and/or uniaxial strain \cite{KanodaKato,RPP}. Therefore, it is possible that the Heisenberg model misses some essential physics as it represents only the lowest order terms in an expansion in $t/U$, where $t$ is the hopping integral between neighbouring sites and the Hubbard $U$ is the effective Coulomb repulsion between two electrons on the same site. The next non-trivial term in this expansion introduces ring exchange into the Hamiltonian \cite{RPP,Montrunich}. It has been argued argued \cite{Montrunich}, in the context of the SL $\kappa$-(BEDT-TTF)$_2$Cu(CN)$_3$ ($\kappa$-CN), that ring exchange and other higher order terms can drive  the breakdown of the 120$^\textrm{o}$ order found on the isotropic triangular lattice before $t/U$ becomes large enough for the Mott transition to occur. Thus, it is clear that the ratio $t/U$ is vitally important in organic charge transfer salts. This has stimulated several groups to attempt to calculate the $U$ from first principles. However, different methods give somewhat inconsistent results \cite{Valenti,Imada,Edan1,Edan2} and so the ratio $t/U$ is not reliably known at present \cite{RPP}. Nevertheless different methods give reasonably consistent trends in the the variation of $U$ across trends. Therefore, we have also investigated the trend in the $U$ associated with [\dmitn]$_2$ dimers. We find very little variation of $U$ across the series, which suggests that ring exchange and other higher-order effects are not \emph{primarily} responsible for different physics observed in the different \dmit salts.

A major impediment to comparing the many body theories described above to experiment has been  the lack of understanding how changing the cation, i.e., choosing $Pn$ and $n$, changes the parameters in the effective Hamiltonians of the monomer and dimer models. The only previous parameterizations of the band structures of the Pd(dmit)$_2$ salts have come from the extended H\"uckel model (a parameterized tight-binding model). However, it has been discovered, e.g., from studies of BEDT-TTF salts \cite{Valenti,Imada,Edan1,Edan2,Pratt}, that the extended H\"uckel model does \emph{not} provide parameters that are accurate enough for discussions of the subtle effects of quantum frustration, which are at play in both the BEDT-TTF and Pd(dmit)$_2$ salts \cite{KanodaKato,RPP}.
Therefore, we performed DFT band structure calculations  in Quantum Espresso \cite{Espresso} using the PBE functional and ultrasoft pseudopotentials with a plane-wave cut-off of 25 Ry a 250 Ry integration mesh. Crystal structures (Fig. \ref{fig:lattice}) are taken to be those observed by x-ray crystallography \cite{KatoJACS,Katoprivate} with only the position of the hydrogen atoms in the cations (which are not visible to x-rays) relaxed. 

In molecular acceptors, such as \dmitn, the Hubbard $U$ associated with a dimer may be written as the sum of two terms $U=U^{(v)}-\delta U^{(p)}$, where $U^{(v)}=E(0)+E(-2)-2E(-1)$ is the $U$ of the dimer {\it in vacuo}, $E(q)$ is the ground state energy of the dimer with charge $q$ and $\delta U^{(p)}$ is the correction due to the polarizable environment of the molecular solid \cite{Edan2}. We  performed DFT calculations $E(q)$, and hence $U^{(v)}$, in SIESTA \cite{Siesta} using a triplet-zeta plus polarization basis set \cite{Edan2}. However, for solids formed from narrow molecules  such as \dmit and BEDT-TTF the accurate calculation $\delta U^{(p)}$ remains an challenging problem \cite{Edan2,Laura}. Therefore, we report $U^{(v)}$, which allows us to understand the trends $U$ across the systems discussed below.

\begin{figure}
\begin{center}
\includegraphics[width=\columnwidth]{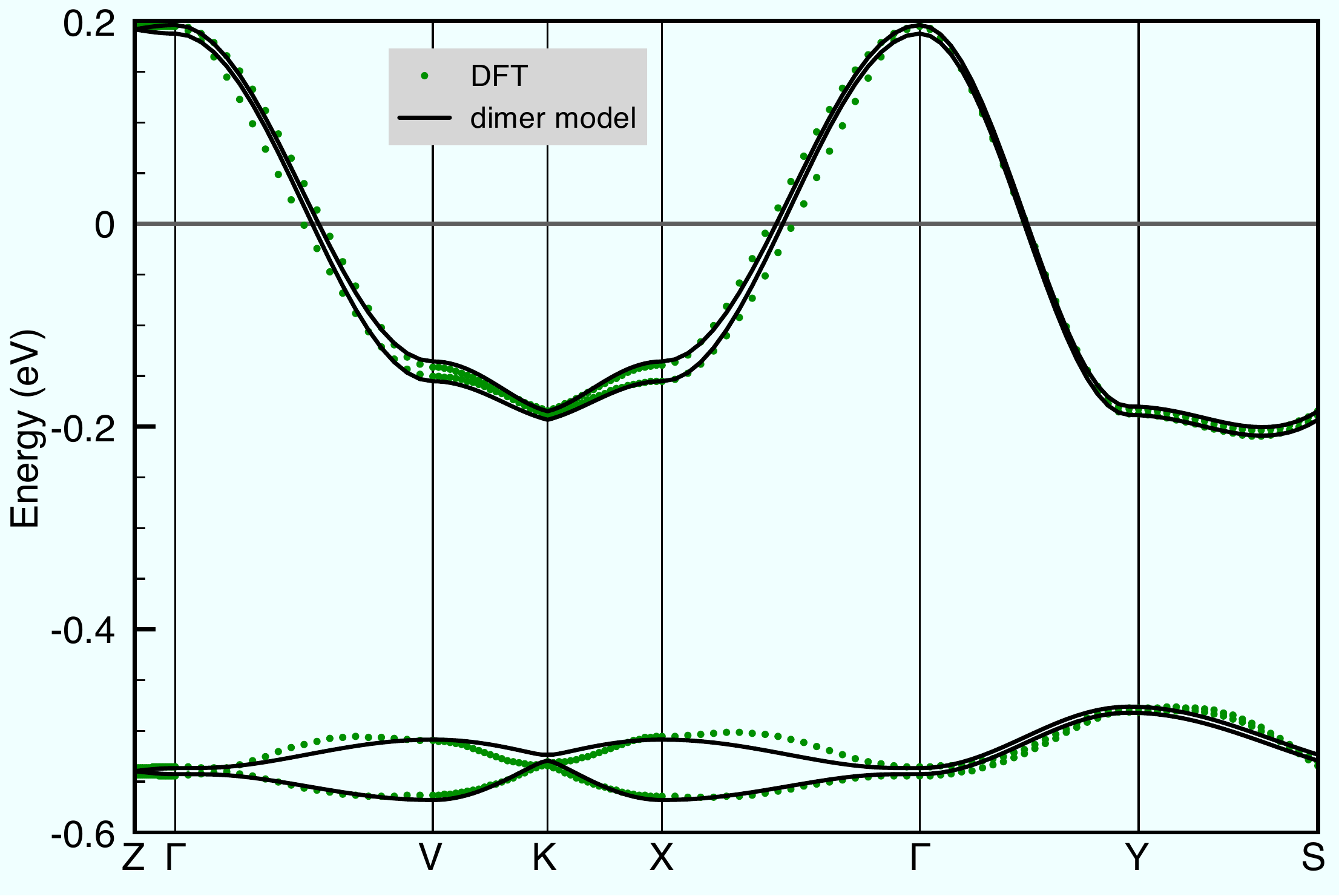}
\end{center}
\caption{(Color online.) Band structure of Sb-1 calculated from DFT (points) and fit to the dimer tight-binding model (curves). The corresponding Fermi surface and the locations of high symmetry points are shown in Fig. 5 of the supp. info. The values of the dimer tight-binding fit (in meV)  for the HOMO band are: $t_{1}=37$, $t_{2}=49$, $t_{3}=45$, $t_{\perp}=-2.1$, 
and the parameters for the LUMO band are $t_{1}=7.2$, $t_{2}=-15$, $t_{3}=-0.2$, $t_{\perp}=-1.5$
. } 
\label{fig:Sb1-bandstructure}
\end{figure}

\begin{figure}
\begin{center}
\includegraphics[width=\columnwidth]{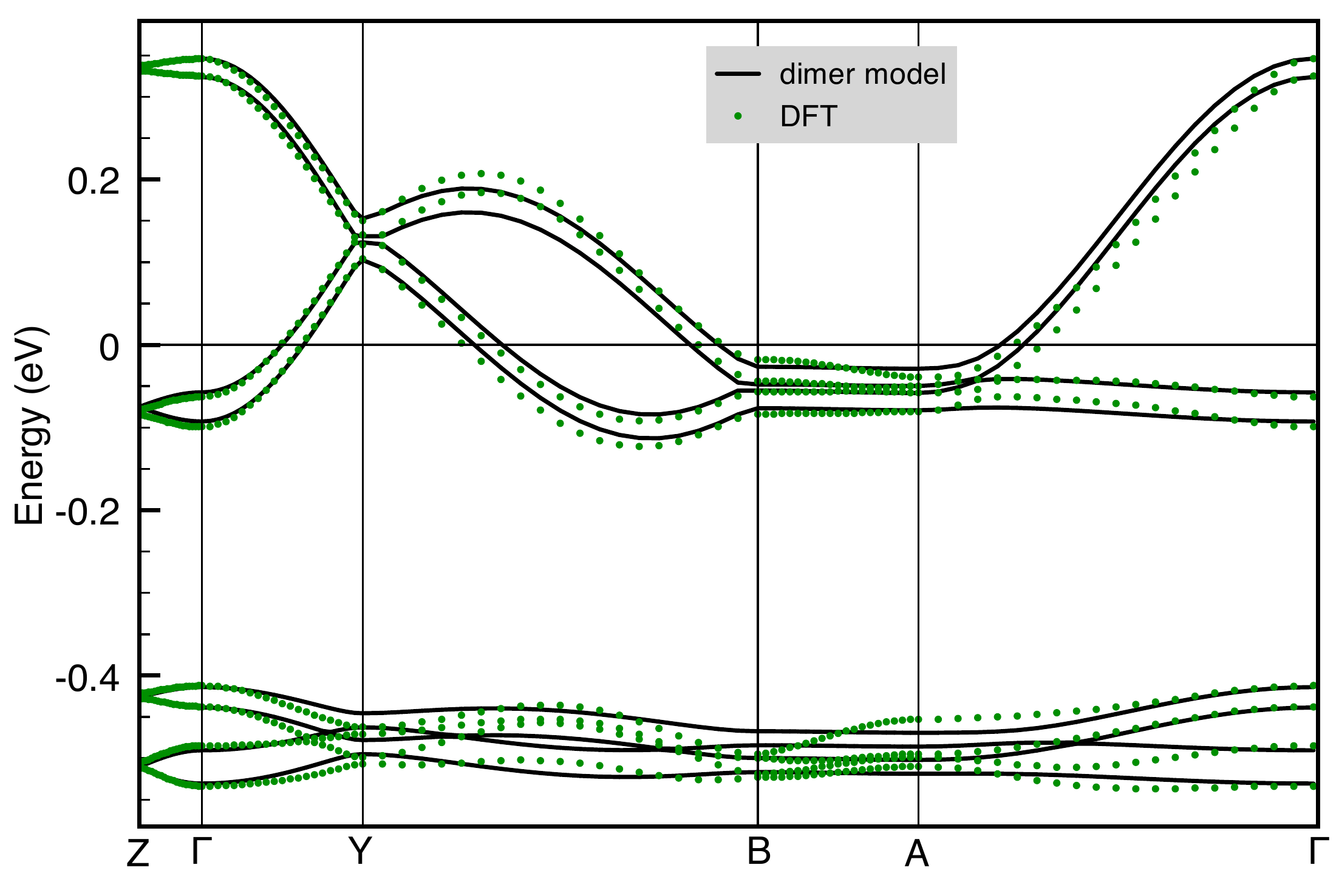}
\end{center}
\caption{(Color online.) Band structure of P-1 calculated from DFT (points) and fit to the dimer tight-binding model (curves). The corresponding Fermi surface and the locations of high symmetry points are shown in Fig. 6 of the supp. info. The values of the dimer tight-binding  fit (in meV)  for the HOMO band are: $t_{1}=45$, $t_{2}=52$, $t_{3}=52$, $t_{\perp}=6.7$, 
$\delta=16$, and the fit to the LUMO band yields $t_{1}=6.6$, $t_{2}=-11$, $t_{3}=-8$, $t_{\perp}=-8.0$, 
$\delta=0$. 
} \label{fig:P1-bandstructure}
\end{figure}

\begin{table*}
\begin{center}
\begin{tabular}{c|ccccccc} 
\hline
Material & expt. & $t_1$ (meV) & $t_2$ (meV) & $t_3$ (meV) & $t'/t$ & $J'/J=(t'/t)^2$ & $U^{(v)}$ [eV] \\ 
\hline 
N-3 & AFM & 54 & 55 & 18 & 0.33-0.33 & 0.10-0.11 & 3.13\\
P-2 & AFM & 68 & 50 & 80 & 0.62-0.73 & 0.38-0.53 & 3.13\\
Sb-2 & CO & 33 & 49 & 46 & 0.69-0.75 & 0.48-0.56 & 3.18\\
Sb-1 & SL & 38 & 49 & 46 & 0.76-0.82 & 0.58-0.67 & 3.12\\
$\kappa$-CN & SL & 43 & 51 & 51 & 0.84 & 0.71 & 2.95 \cite{Edan2}\\
P-1 & VBS & 45 & 51 & 52 & 0.88-0.89 & 0.77-0.80 & 3.12\\
Cs & CO & 42 & 45 & 45 & 0.93 & 0.87 & 3.10\\
Sb-0 & AFM & 38 & 51 & 44 & 1.17-1.29 & 1.36-1.66 & 3.14\\
As-2 & AFM & 39 & 51 & 43 & 1.20-1.31 & 1.44-1.72 & 3.11\\
\hline 
\end{tabular}
\end{center}
\caption{Calculated parameters for the anisotropic triangular lattice for a range of Me$_{4-n}$Et$_nPn$[Pd(dmit)$_2$]$_2$ ($Pn$-$n$) and $\kappa$-(BEDT-TTF)$_2$Cu(CN)$_3$ ($\kappa$-CN). Where two values are given this indicates the small differences because the two $t$ hopping integrals are not identical. Values of $t_1$, $t_2$ and $t_3$ for all compounds are given in the supplementary information; where we also report equivalent results for the local density approximation, which give the same  picture as the PBE results. The column labelled expt. summarizes the experimentally observed low temperature physics. $U^{(v)}$ is the calculated effective Coulomb repulsion between to holes on the same dimer {\it in vacuo}. The variation is small across all of the \dmit salts and even between the \dmit salts and the BEDT-TTF salt. 
}
\label{tab:params}
\end{table*}

Although the in-plane crystal and electronic structures of Sb-1 and P-1 are very similar (cf. Fig. \ref{fig:lattice}), there are some subtle differences that should be noted before we discuss our results in detail. 
Sb-1 takes the, bi-layer, $\beta'$ phase, found in many Pd(dmit)$_2$ salts. In the layer shown molecules stack face-to-face along  $\bf a$-$\bf b$, but in alternate layers molecules stack in the $\bf a$+$\bf b$ direction, where $\bf a$ and $\bf b$ are the crystallographic axes of the conventional unit shown in the Fig \ref{fig:lattice}. The Wigner-Sietz unit cell of Sb-1 is half of this size and therefore only contains one dimer per layer per unit cell. P-1 also forms a bi-layer structure, however, in \dmit molecules stack along the $\bf a$+$\bf c$ axes in both layers. Further, in P-1, the conventional and Wigner-Seitz unit cells are identical and contain two dimers per layer per unit cell. Thus the Wigner-Seitz unit cell of P-1 is approximately twice the size of that of Sb-1 and contains twice as many molecules.

In Fig. \ref{fig:Sb1-bandstructure} we report the calculated band structure of Sb-1. The electronic structure is rather similar to those of N-0 and P-0 \cite{Miyazaki}, which are the only other DFT band structures for salts of \dmit that we are aware of. In all three salts two bands cross the Fermi energy. These bands are derived predominately from the antibonding combination of  highest occupied molecular orbitals (HOMOs). It is not immediately obvious that the HOMOs should be partially filled as \dmit is electron acceptor, one might expect a partial occupation of the lowest unoccupied molecular orbitals (LUMOs). However, the strong hybridisation between the two molecules in each dimer (ringed in Fig. \ref{fig:lattice}) means that the bonding combination of LUMOs is lower in energy than the antibonding combination of HOMOs \cite{RPP,Miyazaki}. There are two bands derived predominately from the antibonding combination of HOMOs and two bands derived predominately from the bonding combination of LUMOs because of the two dimers per unit cell.

In Fig. \ref{fig:P1-bandstructure} we report the calculated band structure of P-1. There are twice as many bands as there are in Fig. \ref{fig:Sb1-bandstructure} because the Winger-Seitz unit cell of P-1 contains four dimers, rather than two. The charge densities corresponding to the bands that cross the Fermi energy in both Sb-1 and P-1 are reported in the Sup. Info. 
We also show fits to the dimer models in Figs. \ref{fig:Sb1-bandstructure} and \ref{fig:P1-bandstructure}. The in plane tight binding integrals are marked in Fig. \ref{fig:lattice}. Interlayer hopping  is described by the hopping integral, $t_{\perp}$. For P-1  we also introduce a parameter $\delta$, which describes the orbital energy differences in the HOMOs (LUMOs) due to the crystallographically distinct local environments of the two dimers per unit cell per layer. Li \etal \cite{Li} have argued for a quarter filled model where each site is a single Pd(dmit)$_2$ molecule. 
Fits to the monomer models are reported in the Sup. Info. Both models provide a good description of the DFT results for both compounds. However, the monomer fit does not appear significantly better than the dimer fit. Therefore we conclude that, at the  level of band structure, the dimer model is sufficient to describe these salts.  

In both materials $t_{2}\simeq t_{3}>t_{1}$. Therefore, in order to make connection with theories of the Heisenberg model on the anisotropic triangular lattice we make the identification $t=\frac12(t_{2}+t_{3})$, $t'=t_{1}$. Thus we find that $t'/t=0.79$ for Sb-1 and $t'/t=0.87$ for P-1.   $J'/J=(t'/t)^2$ to leading order in $t/U$. This yields $J'/J=0.62$ for Sb-1 and $J'/J=0.75$ for P-1. Both of these values are in the regime where the ground state of the Heisenberg model remain controversial. 

We  also calculated the band structures of a range of other salts of \dmitn. The values of $J'/J$ determined analogously to those for Sb-1 and P-1 are reported in Table \ref{tab:params}. Full details of the calculations will be reported elsewhere. We also report our parameterization of the band structure of the $\kappa$-(BEDT-TTF)$_2$Cu(CN)$_3$ ($\kappa$-CN), which has a SL ground state. Note that the value of $t'/t$ for $\kappa$-CN is in excellent agreement with other estimates from DFT (0.83 \cite{Valenti} and 0.80 \cite{Imada}). We also report the calculated values of $U^{(v)}$ for dimers of \dmit in each of these systems in Table \ref{tab:params}. There is no significant variation in $U^{(v)}$ amongst the \dmit salts and even the BEDT-TTF salt has a remarkably similar $U^{(v)}$. The bandwidths $W$ of salts are also remarkably consistent, cf. the Supp. Info., with the  exception of  P-2, which has a rather wider band; however, this material is displays long-range antiferromagnetism. Therefore, we find no evidence that ring exchange is the primary determinant of the magnetic ground state in these materials. This conclusions is supported by experiment. None of the AFM states have been observed to give way to spin-liquid, valence bond solid or charge order when pressure is to them, which drives the materials towards the Mott transition by decreasing $U/t$ and hence increasing the relative importance of ring exchange \cite{KanodaKato,RPP}.  However, note that this comparison of the relative importance of ring-exchange across the series does not imply that ring exchange does not play any role in determining the ground state in the region where there are multiple competing ground states, as we will discuss below.

It is clear from Table \ref{tab:params} that those materials which display long range magnetic order lie in the parameter ranges $J'/J\lesssim0.5$ or $J'/J\gtrsim1$, while those materials which are found experimentally to have CO, SL and VBS ground states all lie in the parameter range $0.5\lesssim J'/J\lesssim0.9$. This is precisely the parameter regime where the many-body ground state remains controversial. This suggests that, in this parameter regime, there are a number of competing ground states, and that other interactions (not included in the anisotropic triangular lattice Heisenberg model) may be important for determining the ground state. Indeed series expansions calculations \cite{series} find a number of ground states with very similar energies in this parameter regime, supporting our contention. Similarly the energies of different phases are found to be similar in this regime in coupled cluster calculations \cite{Bishop}.

Since we placed this manuscript on the arXiv, Hauke \cite{Hauke} has used the compared the tight binding parameters reported in Table \ref{tab:params} to  his self-consistent spin-wave calculations for the Heisenberg model on the completely anisotropic triangular with three distinct exchange interactions. Taking $J_2/J_1=(t_2/t_1)^2$ and  $J_3/J_1=(t_3/t_1)^2$ he finds that our parameters for the completely anisotropic lattice still predict ground states consistent with those found experimentally.

We therefore conclude that the Heisenberg model gives a clear prediction of when long-range magnetic order will be found in the \dmit salts. For $0.5\gtrsim J'/J$ and $J'/J\gtrsim0.9$ the Heisenberg model has a very stable ground state with long range magnetic order (as witnessed by the good agreement between different quantum-many body theories). Thus, perturbations do not change  the nature of the ground state and long range order is realised. But, for 
$0.5\lesssim J'/J\lesssim0.9$  there are a number of  states that are close in energy (consistent with the disagreement in the ground state predicted by different methods). Thus other terms in the Hamiltonian, such as intradimer dynamics, ring exchange, elastic forces in the crystal or the differences in the crystal structures will play an important role in determining which phases are realised. In this light it is interesting to note that the LOMO bands are within $\sim U$ of the HOMO bands, which suggests that they could play an important role. To that end we report the tight binding fits of these bands in Figs. \ref{fig:Sb1-bandstructure} and \ref{fig:P1-bandstructure}. This is in marked contrast from  proposals that proximity to a putative quantum critical point is the determining factor in these systems \cite{ShimizuJPCM,Sachdev}.

We thank Reizo Kato and Ross McKenzie for helpful discussions. This work was supported by the Australian Research Council under the Discovery (DP1093224) and Queen Elizabeth II (DP0878523) schemes and by an award under the MAS on the NCI-NF.


\begin{thebibliography}{99}
\bibitem{defs}
Here Me=methyl, Et=ethyl, $Pn$ is a pnictogen and dmit is 1,3-dithiol-2-thione-4,5-dithiolate.
\end{thebibliography}
\end{document}


\title{Supplimentary Information for ``Geometrical frustration in the spin liquid $\beta'$-Me$_3$EtSb[Pd(dmit)$_2$]$_2$ and the valence bond solid $\beta'$-Me$_3$EtP[Pd(dmit)$_2$]$_2$"}
\author{E. P. Scriven}
\affiliation{Centre for Organic Photonics \& Electronics, School of Mathematics \& Physics,
University of Queensland, Brisbane, Queensland 4072,
Australia}
\author{B. J. Powell}
\email{bjpowell@gmail.com}
\affiliation{Centre for Organic Photonics \& Electronics, School of Mathematics \& Physics,
University of Queensland, Brisbane, Queensland 4072,
Australia}

\pacs{}

\maketitle

As well as the Figs. contained in this document ten movies are also included in the supplementary information for this Letter. These movies allow for three dimensional viewing of the electronic densities of the bands that cross the Fermi levels in P-1 and Sb-1. \url{http://www.youtube.com/watch?v=p2ziIq3WLS0&feature=plcp}, \url{http://www.youtube.com/watch?v=L0cpqNZFAJg&feature=plcp}, \url{http://www.youtube.com/watch?v=6dJpaNWfHks&feature=plcp}, and \url{http://www.youtube.com/watch?v=7_F8iNR3SF0&feature=plcp} show the bands from P-1, and  \url{http://www.youtube.com/watch?v=BaN7WV3BNos&feature=plcp} shows the density summed over these for bands.  \url{http://www.youtube.com/watch?v=GRXouylwTTg&feature=plcp} is a concatenation of all five of  movies for P-1. \url{http://youtu.be/-oER-waMT0E} and \url{http://youtu.be/4d3UAjR3zMo} show the bands from P-1, and  \url{http://youtu.be/cFkfOlOcSOc} shows the density summed over these for bands.  \url{http://youtu.be/92g0xr_fFrk} is a concatenation of all the of these movies for Sb-1.

\begin{figure}[h]
\begin{center}
\includegraphics[width=8cm]{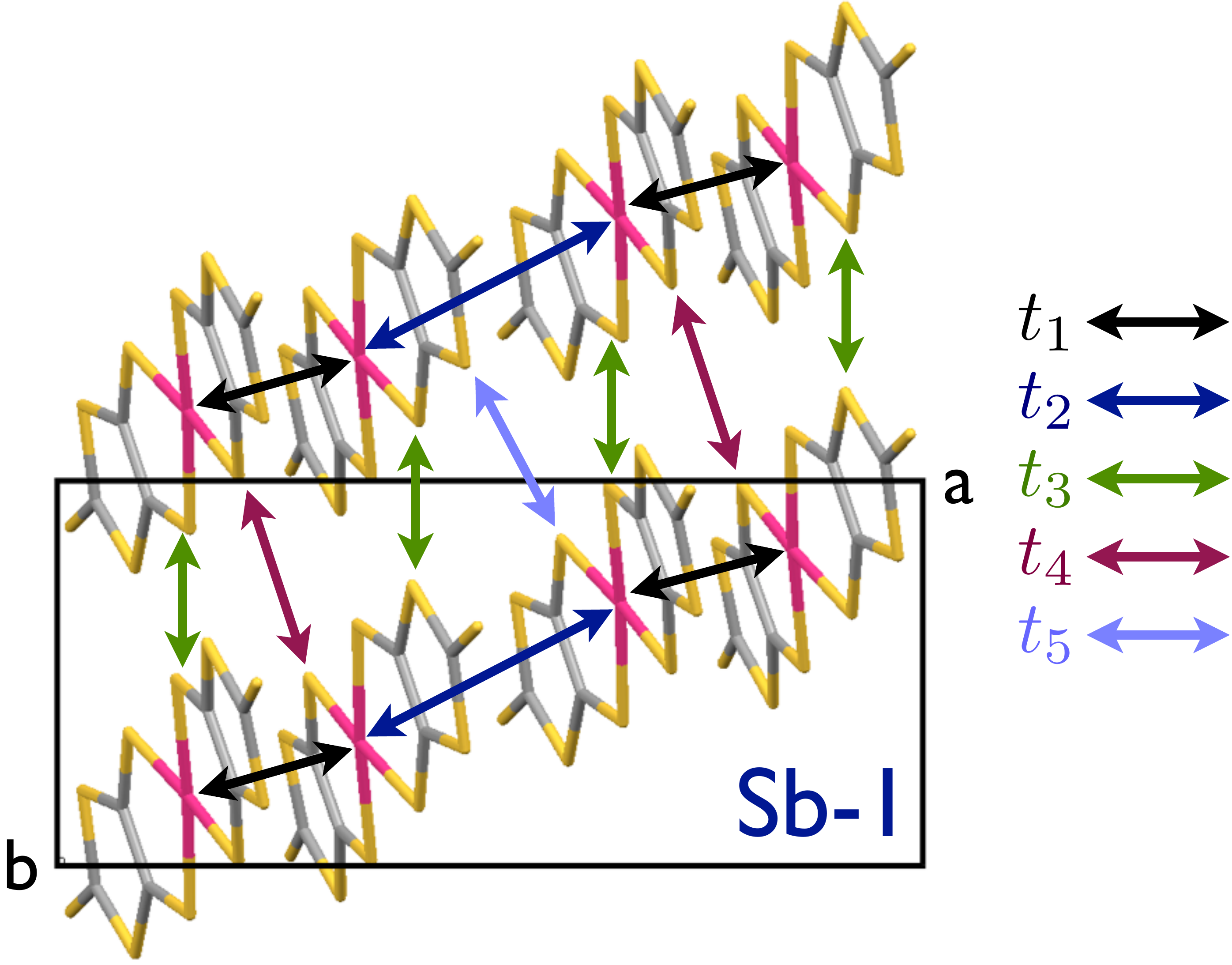}
\end{center}
\caption{The inter-monomer hopping integrals for Sb-1.}\label{fig:lattice3}
\end{figure}

\begin{figure}
\begin{center}
\includegraphics[width=8cm]{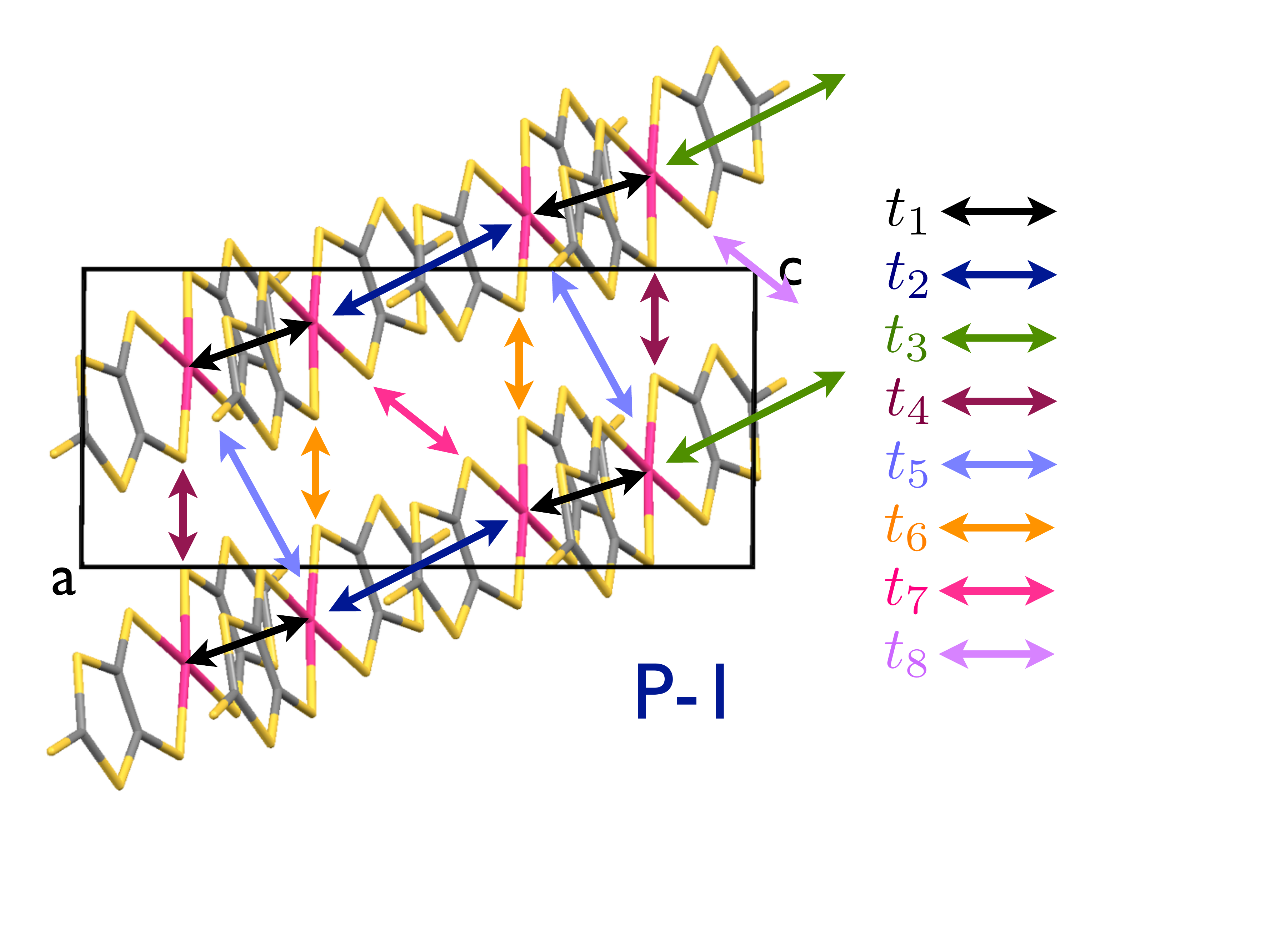}
\end{center}
\caption{The inter-monomer hopping integrals for P-1. 
}\label{fig:lattice4}
\end{figure}

\begin{table*} 
\begin{center}
\begin{tabular}{c|cccccc} 
\hline
Material & expt. & $t_1$ (meV) & $t_2$ (meV) & $t_3$ (meV) & $t'/t$ & $J'/J=(t'/t)^2$ \\ 
\hline 
N-3 & AFM & 54 & 55 & 18 & 0.33-0.33 & 0.10-0.11 \\
P-2 & AFM & 68 & 50 & 80 & 0.62-0.73 & 0.38-0.53 \\
Sb-2 & CO & 33 & 49 & 46 & 0.69-0.75 & 0.48-0.56 \\
Sb-1 & SL & 38 & 49 & 46 & 0.76-0.82 & 0.58-0.67 \\
P-1 & VBS & 45 & 51 & 52 & 0.88-0.89 & 0.77-0.80 \\
Cs & CO & 42 & 45 & 45 & 0.93 & 0.87 \\
Sb-0 & AFM & 38 & 51 & 44 & 1.17-1.29 & 1.36-1.66 \\
As-2 & AFM & 39 & 51 & 43 & 1.20-1.31 & 1.44-1.72 \\
\hline 
\end{tabular}
\end{center}
\caption{Parameters for the bands derived predominately from the HOMOs of the salts of Pd(dmit)$_2$ from fits to the  PBE band structures. Note that as there is no symmetry equivalence between any of the hopping integrals we have taken whichever of $t_1$, $t_2$ or $t_3$ is the most different to be $t'$ - in any case this leads to topologically equivalent aniostropic triangular lattices.}
\label{tab:params}
\end{table*}

\begin{table*} 
\begin{center}
\begin{tabular}{c|cccccc} 
\hline
Material & expt. & $t_1$ (meV) & $t_2$ (meV) & $t_3$ (meV) & $t'/t$ & $J'/J=(t'/t)^2$ \\ 
\hline 
N-3 & AFM & 55 & 55 & 20 & 0.36-0.37 & 0.13 \\
P-2 & AFM & 67 & 45 & 78 & 0.57-0.66 & 0.33-0.44 \\
Sb-2 & CO & 34 & 48 & 45 & 0.71-0.75 & 0.50-0.56 \\
Sb-1 & SL & 37 & 48 & 45 & 0.77-0.83 & 0.60-0.68 \\
P-1 & VBS & 46 & 51 & 52 & 0.89-0.90 & 0.79-0.81 \\
Cs & CO & 48 & 38 & 54 & 0.89-1.29 & 0.80-1.66 \\
Sb-0 & AFM & 39 & 42 & 52 & 1.22-1.32 & 1.17-1.57 \\
As-2 & AFM & 40 & 44 & 51 & 0.87-1.11 & 0.75-1.23 \\
\hline 
\end{tabular}
\end{center}
\caption{Parameters for the bands derived predominately from the HOMOs of the salts of Pd(dmit)$_2$ from fits to the  LDA band structures.}
\label{tab:params2}
\end{table*}

\begin{figure}
\begin{center}
\includegraphics[width=8cm]{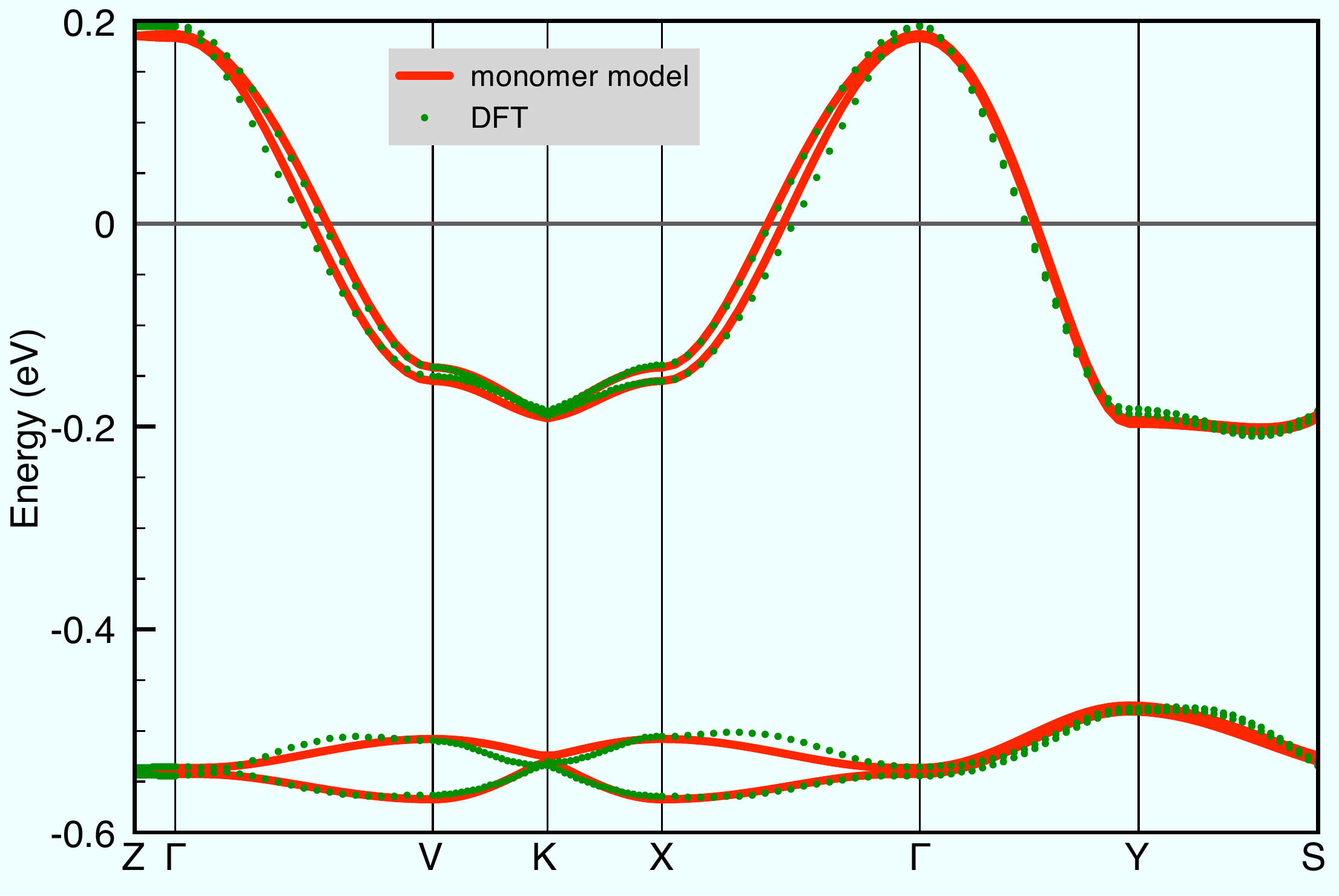}
\end{center}
\caption{(Color online.) Band structure of Sb-1 calculated from DFT (points) and fit to the monomer tight-binding model (curves). The location of high symmetry points are marked in Fig. 4 of the main text.  The values of the monomer tight-binding fit to the HOMO are (in meV): $t_{1H}=426$, $t_{1L}=470$, $t_{2H}=92$, $t_{2L}=0.6$, $t_{3H}=1.8$, $t_{3L}=17$, $t_{4H}=68$, $t_{4L}=19$, $t_{H}=98$, $t_{5L}=30$, $t_{\perp H}=1.0$, $t_{\perp L}=1.5$, $\nu=-427$, $\delta=225$, where the subscript $H$ indicates parameters for the HOMO band and the subscript $L$ indicates parameters for the LUMO band.} 
\label{fig:Sb1-bandstructure}
\end{figure}

\begin{figure}
\begin{center}
\includegraphics[width=8cm]{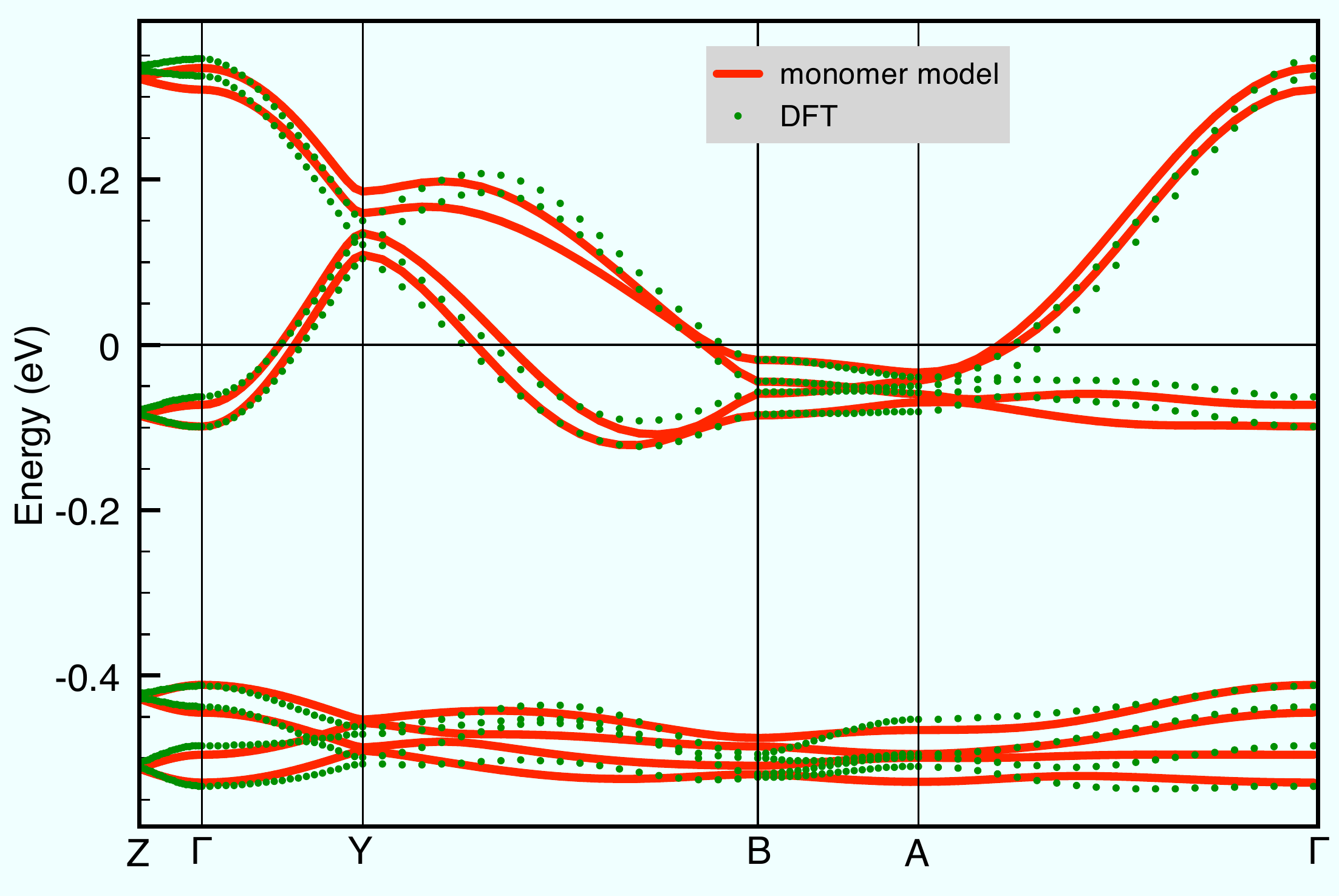}
\end{center}
\caption{(Color online.) Band structure of P-1 calculated from DFT (points) and fit to the monomer tight-binding model (curves). The location of high symmetry points are marked in Fig. 4 of the main text. The values of the monomer tight-binding fit are (in meV): $t_{1H}=603$, $t_{1L}=493$, $t_{2H}=84$, $t_{2L}=32$, $t_{3H}=124$, $t_{3L}=24$, $t_{4H}=-68$, $t_{4L}=5.7$, $t_{5H}=67$, $t_{5L}=15$, $t_{6H}=-87$, $t_{6L}=22$, $t_{7H}=105$, $t_{7L}=13$, $t_{8H}=104$, $t_{8L}=15$, $t_{\perp H}=-6.5$, $t_{\perp L}=8.4$, $\nu=-400$, $\delta=300$, where the subscript $H$ indicates parameters for the HOMO band and the subscript $L$ indicates parameters for the LUMO band.
} \label{fig:P1-bandstructure}
\end{figure}

\begin{figure}
\begin{center}
\includegraphics[width=8cm, angle=0]{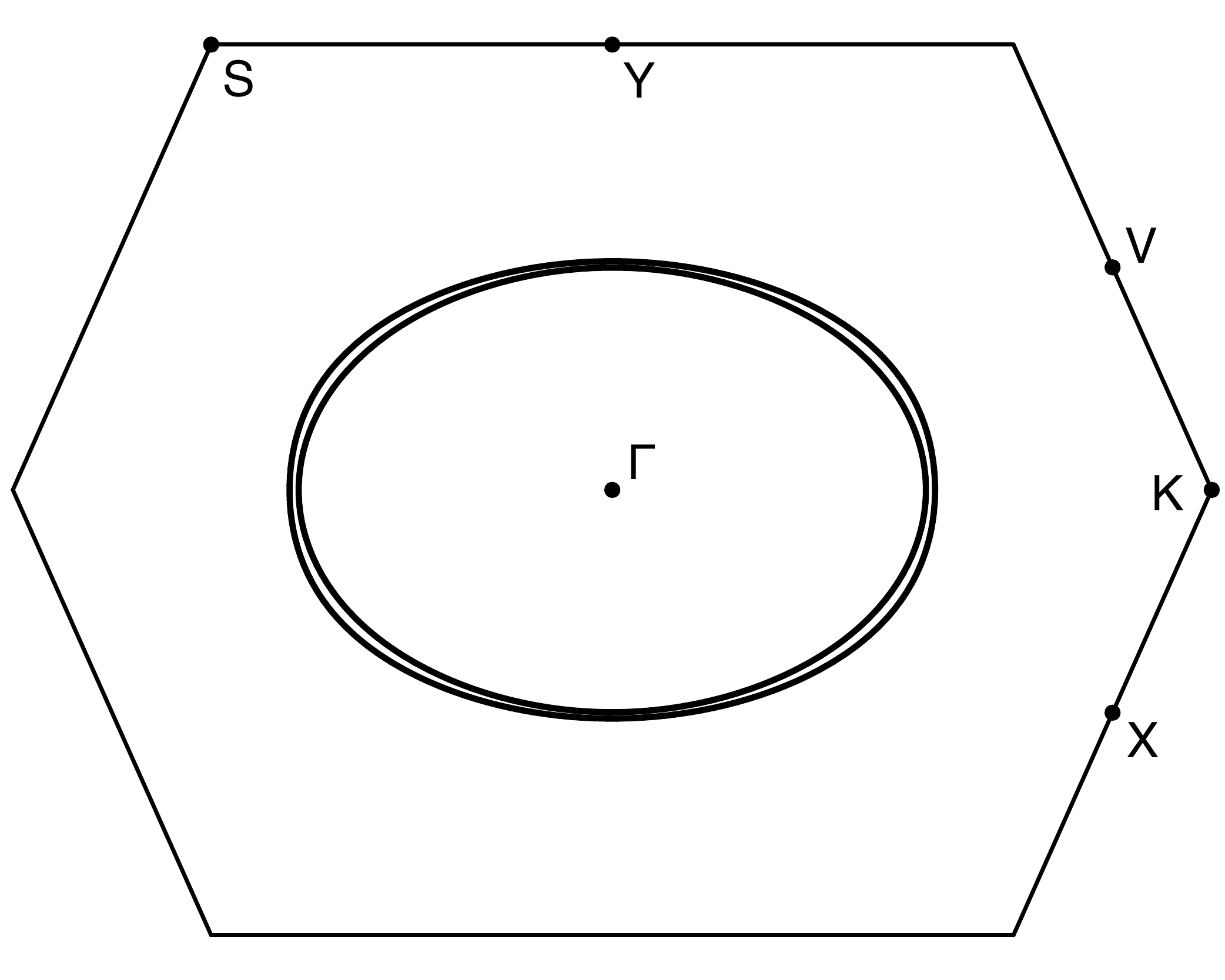}
\end{center}
\caption{Fermi surfaces of Sb-1  calculated from the dimer model.}
\label{fig:FSSb1}
\end{figure}

\begin{figure}
\begin{center}
\includegraphics[width=8cm, angle=0]{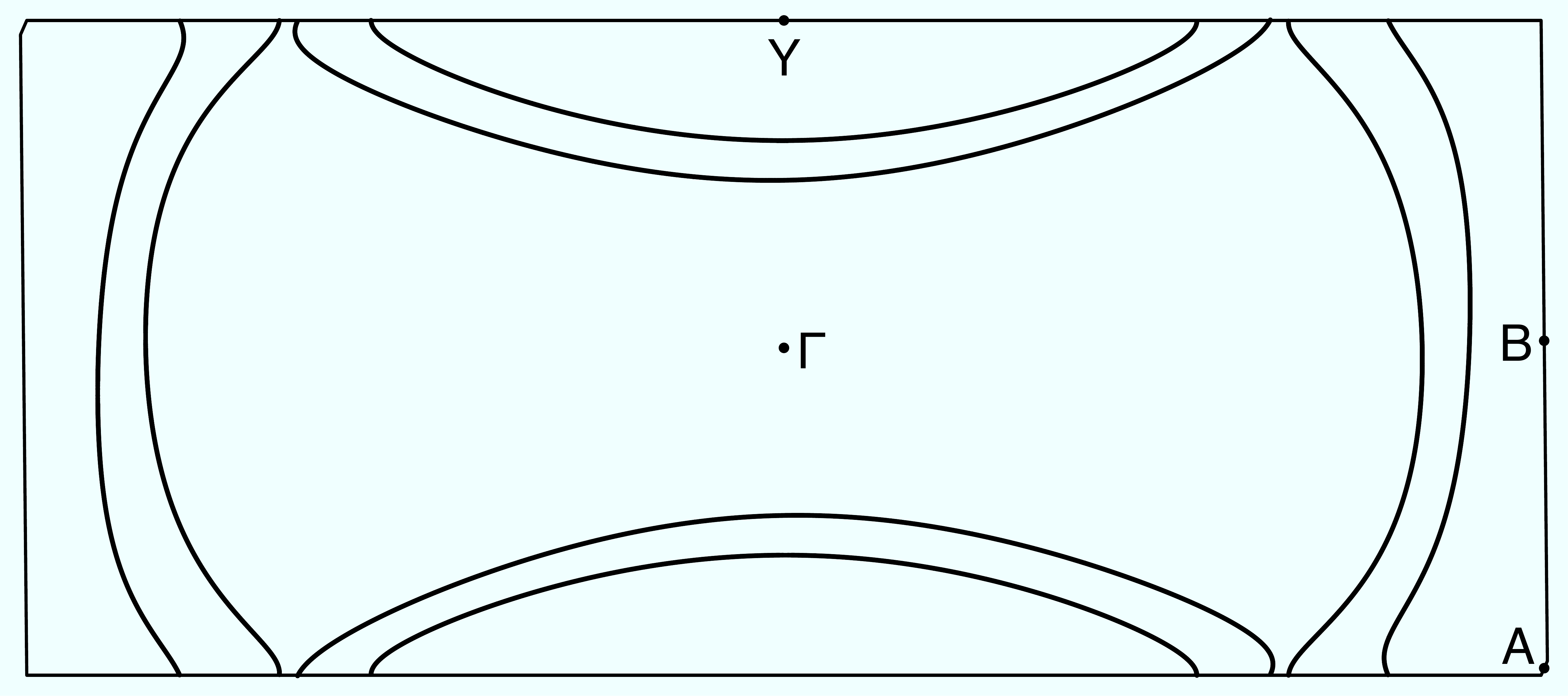}
\end{center}
\caption{Fermi surfaces of P-1  calculated from the dimer model.}
\label{fig:FSP1}
\end{figure}

\begin{figure}
\begin{center}
\includegraphics[width=8cm]{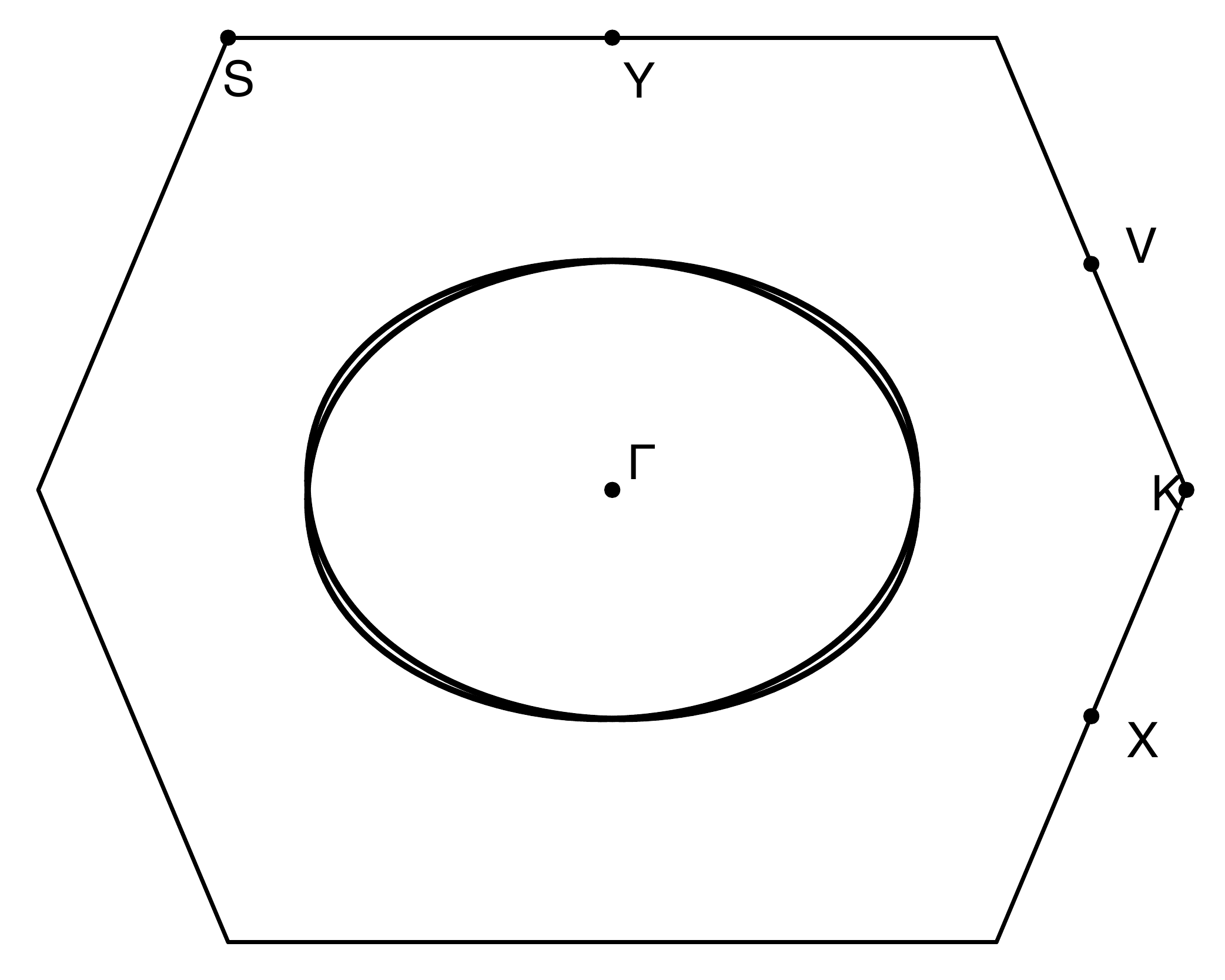}
\end{center}
\caption{Fermi surface of Sb-1 calculated from the tight binding dimer model for the parameters fit to the DFT calculation, but with $t_\perp=0$. Note that now there is no hybridiation between the two bands that cross the Fermi energy.}
\label{fig:FSSb1tperpeq0}
\end{figure}

In Figs. \ref{fig:FSSb1} and \ref{fig:FSP1} we plot the Fermi surfaces of both materials. There are two quasi-two dimensional sheets in the Fermi surface of Sb-1. These arise essentially from the bonding and antibonding combinations of the Fermi surfaces in the two layers. ($t_{\perp}$, which is responsible for the hybridisation, is a perturbation as it is an order of magnitude smaller than the in plane hopping integrals.) Because the \dmit molecules stack along different direction in alternate layers, for $t_{\perp}=0$ one would have two uncoupled approximately elliptical Fermi surface sheets; one `pointing' towards the V and the other towards  X (cf. Fig. \ref{fig:FSSb1} and Fig. \ref{fig:FSSb1tperpeq0}). 
It is interesting to note that both DFT and tight binding models implicitly assume coherent charge transport between the layers. However, a number of metallic organic charge transfer salts do not display coherent interlayer transport \cite{Singleton}. In the metallic phases of these materials, induced by the application of pressure \cite{KanodaKato}, this could lead to rather interesting phenomenology. For example, because of the complicated shape of the first Brillouin zone (FBZ) in Sb-1 the three dimensional Fermi surface occupies less than half of the cross-sectional area of the FBZ in the plane (shown in Figs. \ref{fig:FSSb1} and \ref{fig:FSP1}) despite being half-filled throughout the entire FBZ. Therefore, one might expect a dramatic change in the cross-sectional area observed, for example, in quantum oscillation experiments if interlayer coherence is lost and a true two dimensional Fermi surface, which is half filled in the plane is formed. More subtle restructuring of the Fermi surfaces may also result in either salt from the loss of hybridisation between the multiple sheets of Fermi surface, as this is mediated by $t_\perp$.
The doubling of the unit cell in P-1 (compared to Sb-1) halves  the FBZ. This results in a folding of the Fermi surface - similar to the differences between the Fermi surfaces of the $\beta$ and $\kappa$ phases of BEDT-TTF salts \cite{Jaime-t}. Otherwise the Fermi surfaces of the two materials are remarkably similar.

\begin{figure}
\begin{center}
\includegraphics[width=\columnwidth]{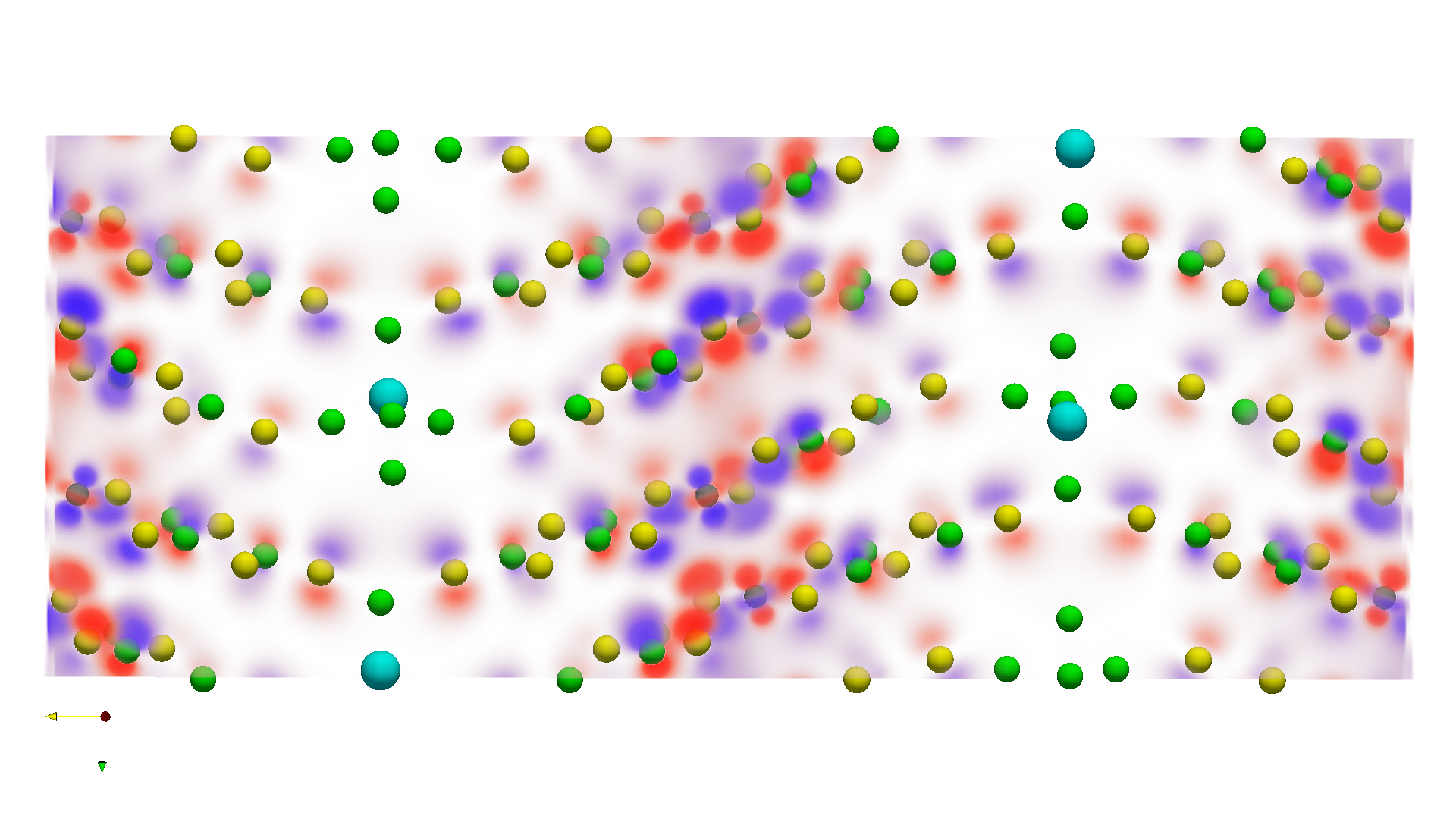}
\end{center}
\caption{Electron density for electrons in band 453 of P-1. Colour represents phase of the wavefunction. Spheres are atomic nuclei. Protons are not shown. A rotating animation of this plot, which allows for better visualisation of the entire orbital is also available on \texttt{youtube} as part of the supplementary information, see main text for link.}
\end{figure}

\begin{figure}
\begin{center}
\includegraphics[width=\columnwidth]{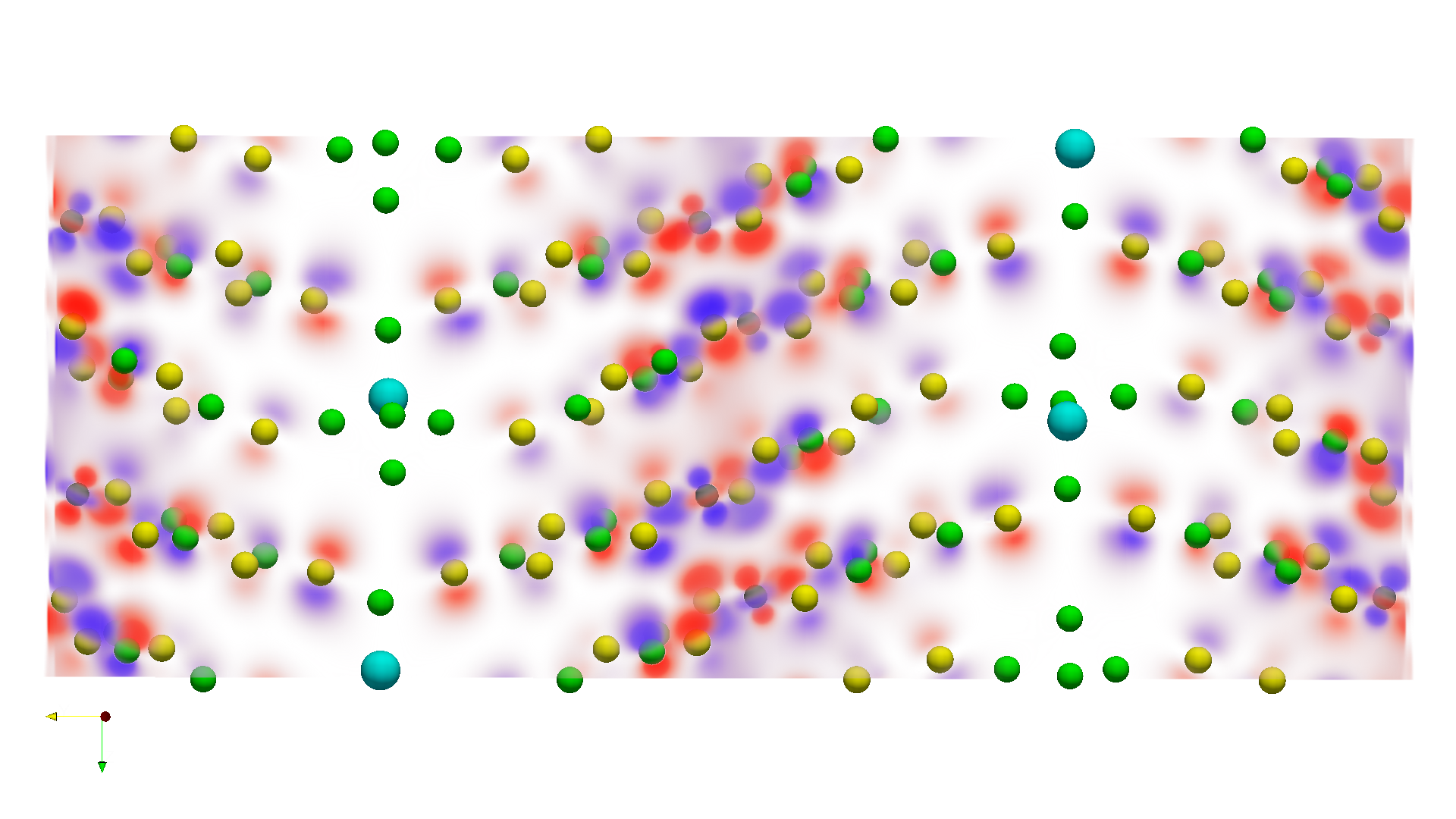}
\end{center}
\caption{Electron density for electrons in band 454 of P-1. Colour represents phase of the wavefunction. Spheres are atomic nuclei. Protons are not shown. A rotating animation of this plot, which allows for better visualisation of the entire orbital is also available on \texttt{youtube} as part of the supplementary information, see main text for link.}
\end{figure}

\begin{figure}
\begin{center}
\includegraphics[width=\columnwidth]{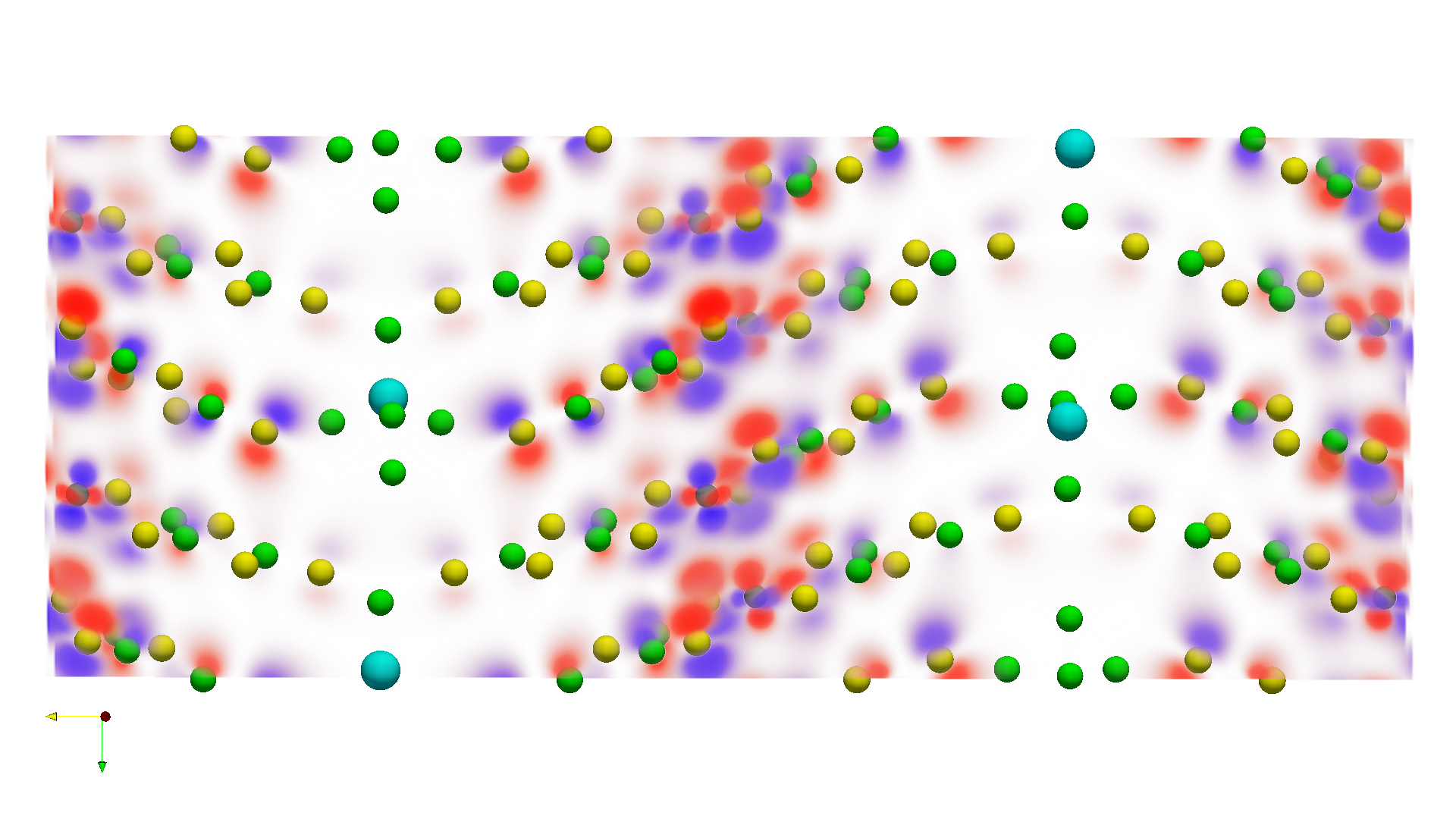}
\end{center}
\caption{Electron density for electrons in band 455 of P-1. Colour represents phase of the wavefunction. Spheres are atomic nuclei. Protons are not shown. A rotating animation of this plot, which allows for better visualisation of the entire orbital is also available on \texttt{youtube} as part of the supplementary information, see main text for link.}
\end{figure}

\begin{figure}
\begin{center}
\includegraphics[width=\columnwidth]{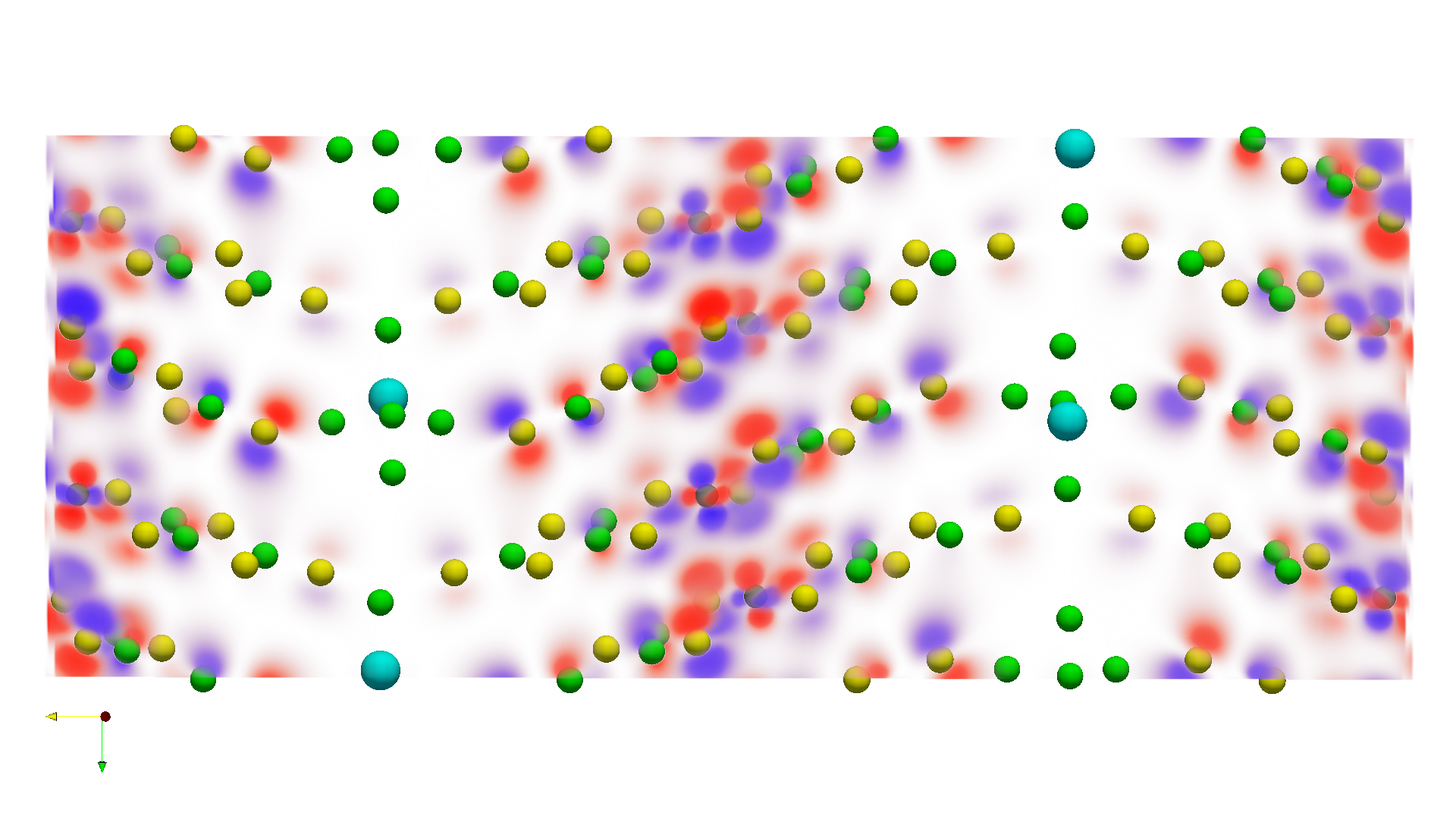}
\end{center}
\caption{Electron density for electrons in band 456  of P-1. Colour represents phase of the wavefunction. Spheres are atomic nuclei. Protons are not shown. A rotating animation of this plot, which allows for better visualisation of the entire orbital is also available on \texttt{youtube} as part of the supplementary information, see main text for link.}
\end{figure}

\begin{figure}
\begin{center}
\includegraphics[width=\columnwidth]{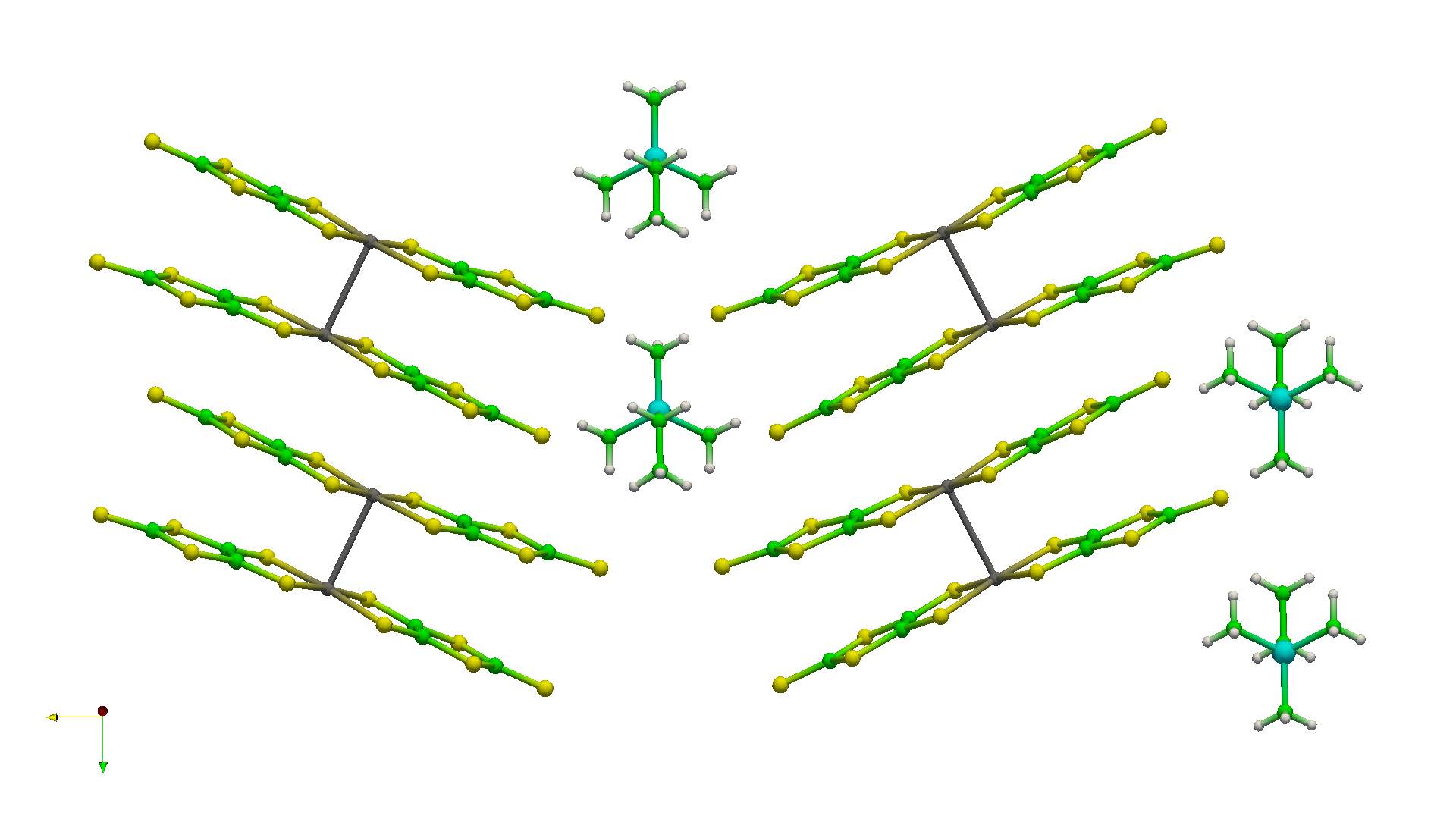}
\end{center}
\caption{Structure of P-1. For comparison with Figs. 6-9.}
\end{figure}

\begin{figure}
\begin{center}
\includegraphics[width=\columnwidth]{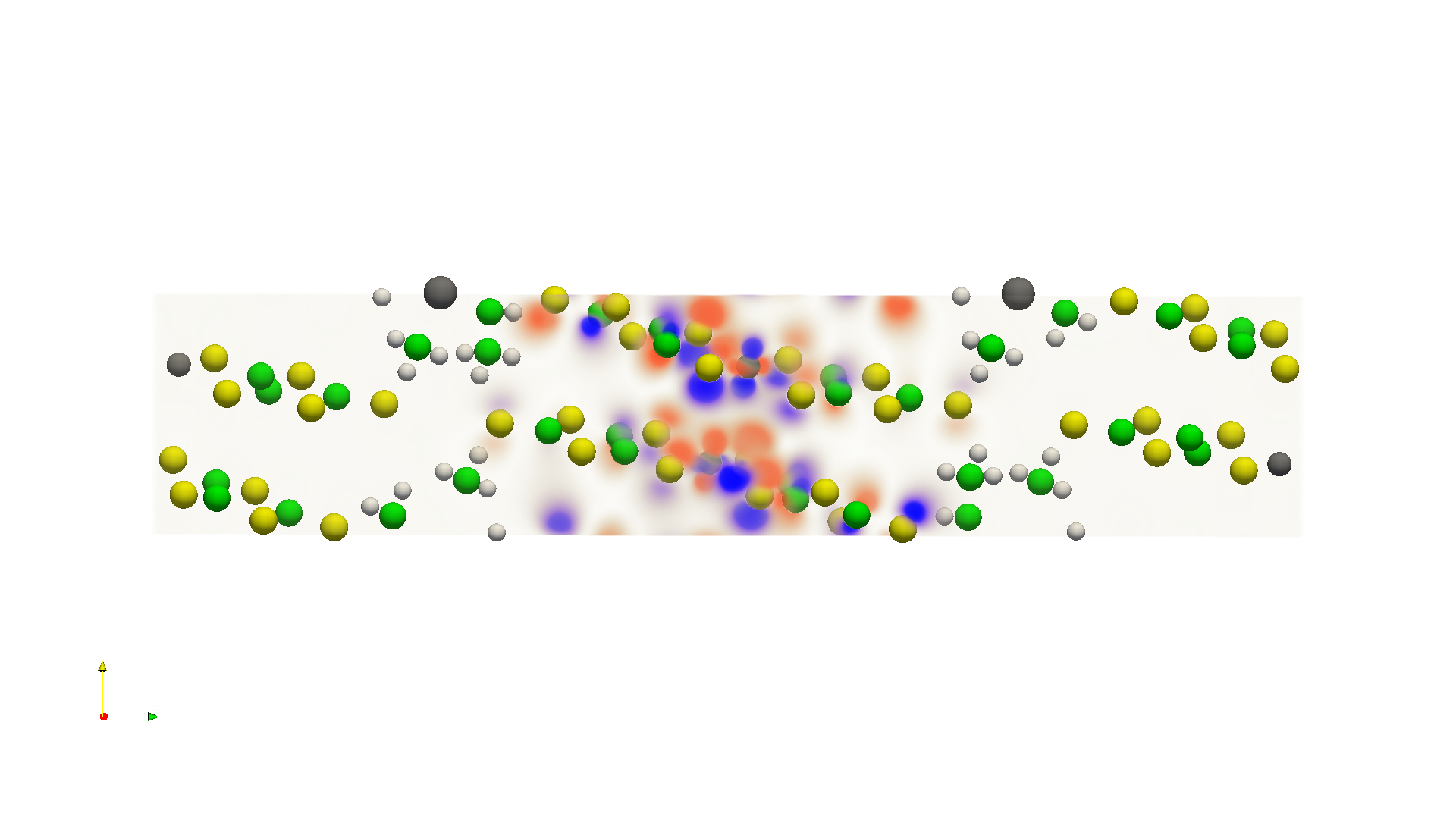}
\end{center}
\caption{Electron density for electrons in band 227 of Sb-1. Colour represents phase of the wavefunction. Spheres are atomic nuclei.  A rotating animation of this plot, which allows for better visualisation of the entire orbital is also available on \texttt{youtube} as part of the supplementary information, see main text for link.}
\end{figure}

\begin{figure}
\begin{center}
\includegraphics[width=\columnwidth]{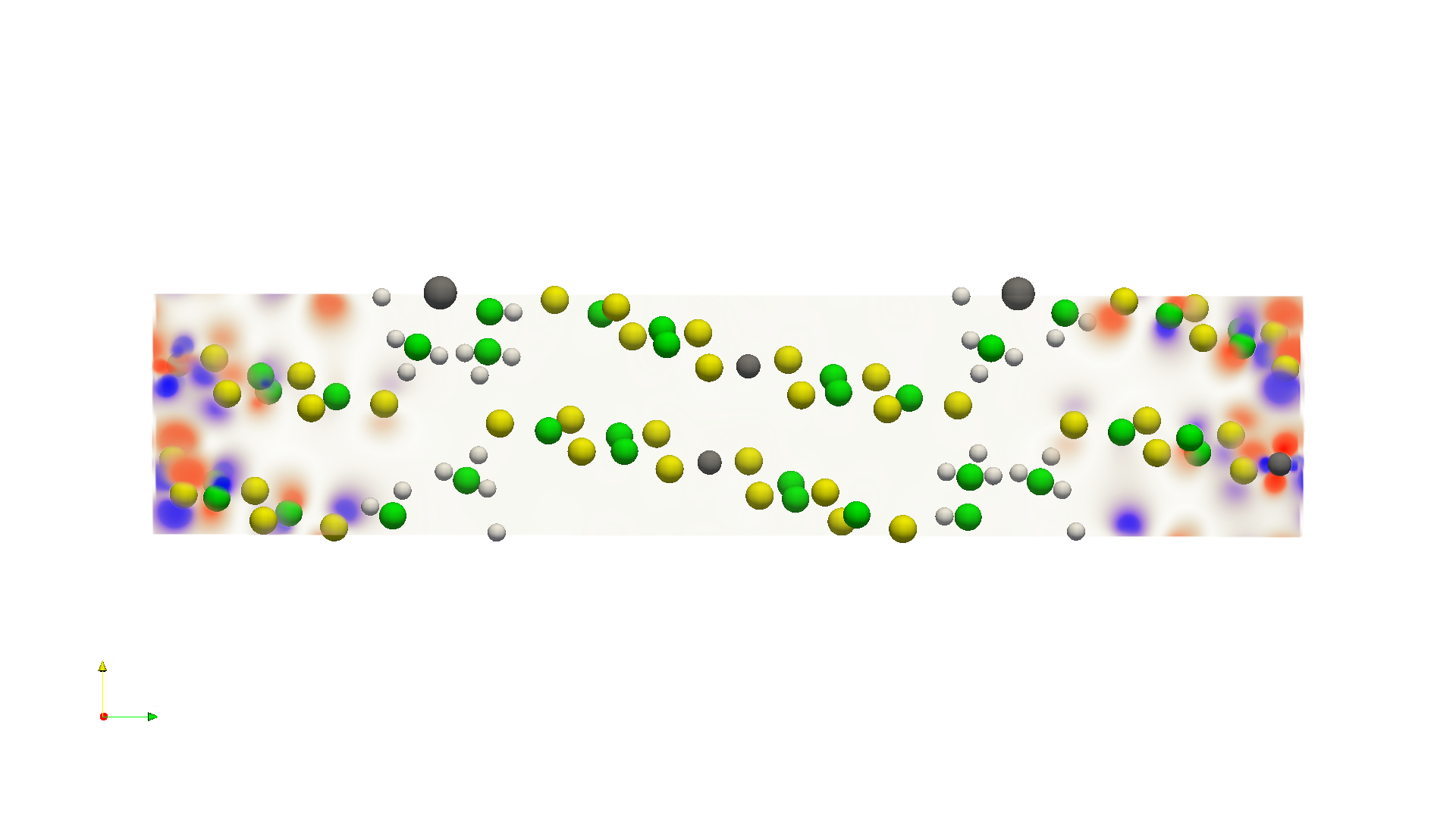}
\end{center}
\caption{Electron density for electrons in band 228  of Sb-1. Colour represents phase of the wavefunction. Spheres are atomic nuclei.   A rotating animation of this plot, which allows for better visualisation of the entire orbital is also available on \texttt{youtube} as part of the supplementary information, see main text for link.}
\end{figure}

\begin{figure}
\begin{center}
\includegraphics[width=\columnwidth]{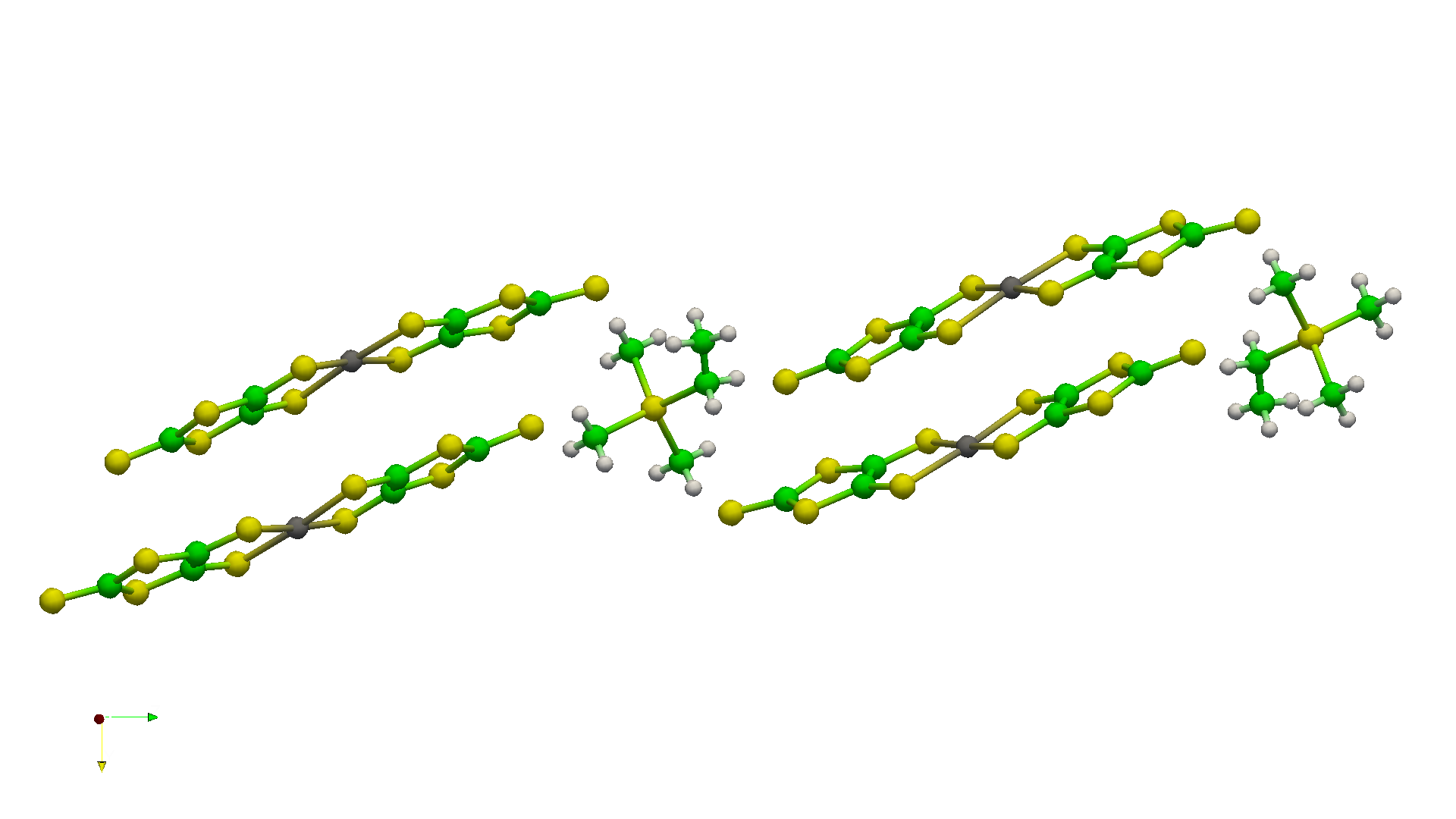}
\end{center}
\caption{Structure of Sb-1. For comparison with Figs. 11 and 12.}
\end{figure}
